\documentclass[twocolumn,showpacs]{revtex4}
\usepackage{wrapfig,epsfig,epsf}
\usepackage{dcolumn}
\usepackage{amssymb,amsmath}
\usepackage{mathtext}
\usepackage{rotate}
\usepackage{euscript}
\usepackage{bm}

\begin{document}

\input epsf
\draft
\renewcommand{\topfraction}{0.8}

\title {\Large\bf The generality of inflation in closed cosmological models with some quintessence 
potentials  }
 \author{\bf S. A. Pavluchenko}
\affiliation{ { Sternberg Astronomical Institute, Moscow State University, Moscow 119992, Russia}    }

{\begin{abstract}
We have investigated the generality of inflation (the probability of inflation in other words) in 
closed FRW models
 for a wide class of quintessence potentials. It is shown that inflation is not suppressed for most of 
 them and for a wide
 enough range of their parameters. It allows us to 
 decide inflation is common enough even in the case of closed Universe. 
\end{abstract}}

\maketitle

\section{Introduction}

Recent observations of the supernovae type Ia~\cite{snIa_1,snIa_2} combined with the CMB 
data~\cite{CMB_rec}
and the data on large scale structure~\cite{LSS_rec} provide us with evidence our Universe is 
accelerating. One can 
explain it via a presence of the small positive $\Lambda$-term (cosmological constant). Here we have a 
deal with one of kinds
of the dynamical $\Lambda$-term namely quintessence (see for review~\cite{als_var00,var02,peeb_ratr02}
). It can explain the stage of inflation expansion~\cite{infl} and acceleration nowadays,
this is the reason for improving interest for it recently. But the theories with the scalar 
field
as a source of expansion have a free parameter~-- the potential of this scalar field. Yes, there were
attempts to restore kind of the potential from SN Ia observations~\cite{als-sn-v}, but they did not 
give us an
answer to our question about an exact kind of the potential. So one have a deal with different potentials 
of the scalar field in different areas of physics, not in the cosmology only. And the aim of this paper is 
the 
test of some of these potentials, attracted an attention recently. Namely we want to study the 
degree of inflationarity in 
Friedman-Robertson-Walker (FRW) models with different scalar field potentials motivated
by particle physics, galaxies rotation curves, etc. 

Saying {\it the degree of inflationarity} we mean the ratio of number of solutions experienced 
inflation
to number of all possible ones. And saying about the number of solutions we mean not exact number~-- 
there is infinite number of all possible solutions~-- but the number one can obtain using evenly 
distributed
net on the space of initial conditions. But choosing this net we are really choosing the {\it measure} on 
the initial conditions space and after doing it we can speak about the {\it probability}~-- very ratio 
is the probability of inflation for our model in the sense of measure we chose. 
So if this ratio is small
enough for some potentials one can decide in the model with this potential the inflation is suppressed.

Here we work with closed models~-- curvature is the parameter which is "forgotten" during inflation,
but before inflation curvature had acted very strongly so presence of nonzero curvature will change the
probability of inflation. By the way, recent CMB data~\cite{CMB_rec} show $\Omega_0 = 1.02 \pm 0.02$
(but such a models~-- closed FRW~-- have some problems~\cite{Linde_rec})
In the case of initially open or flat Universe the scale factor of the Universe $a$ 
cannot pass through extremum points. In this case all the trajectories starting from a sufficiently 
large initial value of the scalar field $\varphi_0$ reach a slow-roll regime and experience inflation.
Here we call as trajectory 
the evolution curve of the Universe in some coordinates 
(for example ($a$,$t$) or as in \cite{we99} ($a$,$f$)).
If we start from the Planck energy a measure of non-inflating trajectories for a scalar field with the
mass $m$ is about $m/m_{P}$. From observational reasons, this ratio is about $10^{-5}$ so almost all 
trajectories lead to the inflationary regime. However, positive spatial curvature allows a trajectory 
to have a point of maximal expansion and results in increasing the measure of non-inflating 
trajectories (\cite{four,B-Kh}).

The structure of the paper is following: first we write down main equations and introduce our measures. 
Then we have a deal with different potentials, motivated by observation and by particle physics and 
attracted an attention last years. For all of them we say some words about their origins with 
corresponding references and give the results for the degree of inflationarity in FRW models filled 
with the scalar field with these potentials.

\section{Main equations}

Here we follow~\cite{we01gr,we01ijmpd}. The equations describing the evolution of the 
Universe in closed FRW model are  
$$
\frac{m_P^2}{16 \pi}\left(\ddot a+\frac{{\dot a}^2}{2a}+\frac{1}{2a} \right) + \frac{a}{4} \left( 
\frac{{\dot \varphi}^2}{2}- V(\varphi) \right) =0,  
$$
$$
\ddot \varphi + \frac{3 \dot \varphi \dot {\vphantom{\varphi}a} }{a}+\frac{dV(\varphi)}{d\varphi} =0, 
$$
\noindent and the first integral of the system is
$$
\frac{3 m_P^2}{8 \pi}\left(\frac{\dot a^2}{a^2} + \frac{1}{a^2}\right)=\left( V(\varphi)+\frac{{\dot 
\varphi}^2}{2} \right). 
$$

\begin{figure*}
\rotatebox{-90}{
\epsfxsize=12cm
\includegraphics{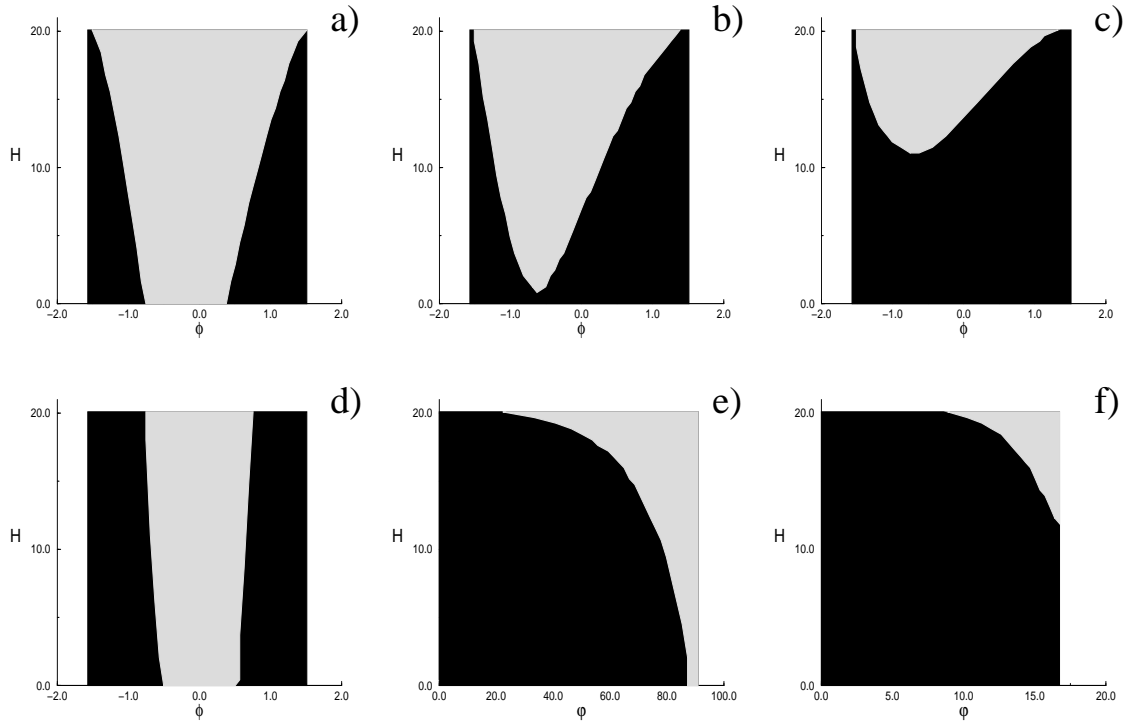}}
\caption
{Examples of the areas on the initial conditions space which lead to inflation (grey dots) and don't 
lead 
(black ones). Figs (a)--(c) correspond to the case of the inverse power-law potential, (d)~-- to the 
exponential potential and (e)--(f)~-- to the FCDM potential (see Sec. III F.)~-- field parametrization 
(see text for details).}
\end{figure*}

Now we need to introduce parametrization~-- it determines the measure we use. 
For the FRW case the most common view of the 
{\it trigonometrical (angular)} parametrization $(\phi,H)$ is
\begin{widetext}
\hfill (*)
\begin{eqnarray*}
\frac{3 m_P^2}{8 \pi} \left( H^2 + \frac{1}{a^2} \right) = m_P^4, \quad 
H^2 + \frac{1}{a^2} = \frac{8 \pi m_P^2}{3}, \quad H \in \left[ 0; \sqrt{\frac{8\pi}{3}}m_P \right); \\
 V(\varphi) + \frac{ {\dot \varphi}^2 }{2} = m_P^4, \quad V(\varphi) = m_P^4 \cos^2 \phi,
\quad \frac{ {\dot \varphi}^2 }{2} = m_P^4 \sin^2 \phi, \quad \phi \in \left[ -\frac{\pi}{2}; \frac{\pi}{2}
\right]. 
\end{eqnarray*}
\end{widetext}

Here we have two dimensionless parameters: $\phi$ and the initial value of Hubble parameter $H$.
This parametrization is very suitable for the potentials like power-law but unapplicable for example in 
the case of the complex scalar field~\cite{we01gr,we01ijmpd}. In~\cite{we01gr,we01ijmpd} we have introduced
another parametrization~-- {\it field} one $(\varphi,H)$. It is
determined as following~-- we will use initial value of the field $\varphi$ instead of $\phi$ and 
second coordinate is the
same~-- initial value of the Hubble parameter $H$. This parametrization also have a disadvantage~-- 
it's unapplicable for example for the 
runaway potentials. 

Also we need to introduce the Planck boundary
$$
V(\varphi) + \frac{ {\dot \varphi}^2 }{2} = m_P^4. 
$$

And our work is following: starting from the Planck boundary for given pair of the initial conditions 
($(\phi,H)$ or $(\varphi,H)$) we
numerically calculate further evolution of the Universe to get an answer~-- will our Universe 
experience
inflation or not. 
Here we don't check 60 e-foldings 
so when the scale factor increase in $10^5$ times, we call it as inflation
(also we are checking for equation of state~-- 
during inflation the scalar field must violate the strong energy condition). 
And doing scans on all possible values
of the initial conditions (with measure used we have bounded area of initial conditions) we get the ratio of
the inflation solutions to all ones~-- the probability of inflation in some sense (one can use both the 
{\it probability} of inflation for the model with given potential and the {\it degree of 
inflationarity} of the model with given potential to call very ratio).

\section{Potentials}

Before telling about the potentials we use, let us note~-- most of analysis' given in references are made
for the case of flat Universe. Well, we are living now in a flat (with good precision) Universe and 
so we have a deal with quintessence in flat Universe. But our Universe is flat due to the inflation 
and before inflation it could be curved. As we already noted above, here we have a deal with closed 
Universe. 

In this paper we will use normalization $m_P^2/16\pi \equiv 1$ and one needs to keep it in mind while 
recalculating all values (like parameters in potentials, etc.) to more common quantities.

\subsection{Power-law potentials}

\begin{figure}
\epsfxsize=9cm
\centerline{{\epsfbox{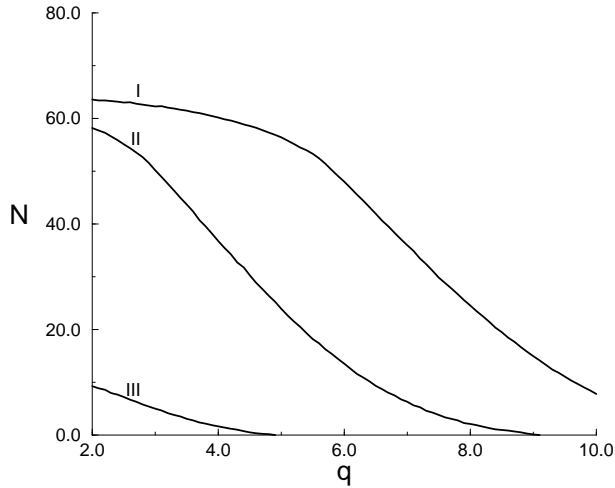}}}
\caption
{Inverse power-law potential: the dependence of the degree of 
inflationarity on the power $q$. I corresponds to the case $M~\sim~m_P $, II~-- $M~\sim~0.7~m_P$ and 
III~-- $M~\sim~0.4~m_P$.}
\end{figure}

\begin{figure}
\epsfxsize=9cm
\centerline{{\epsfbox{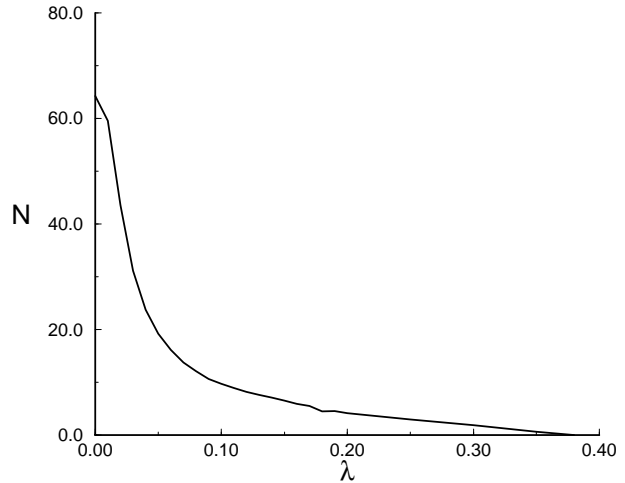}}}
\caption
{Exponential potential: the dependence of the degree of 
inflationarity on $\lambda$. To recalculate $\lambda$ to more common values one needs to keep in mind
$m_P^2/16\pi \equiv 1$.}
\end{figure}

\begin{figure*}
\epsfxsize=12cm
\centerline{{\epsfbox{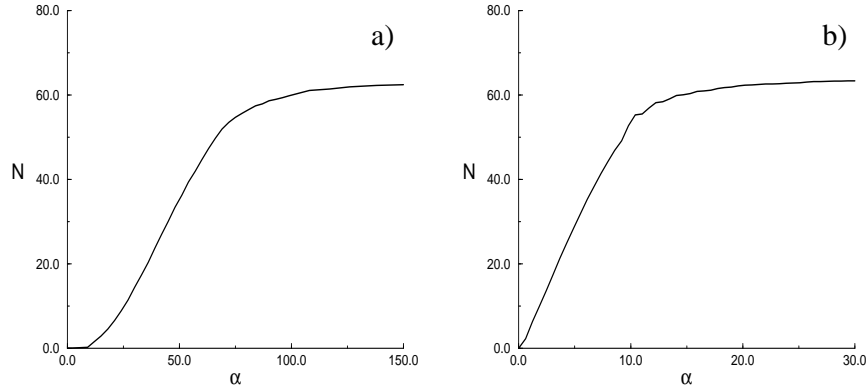}}}
\caption
{ZWS potential: the dependence of the degree of 
inflationarity on $\alpha$ for $V_0~\sim~4~\times~10^{-4}~m_P^4$ (a) and 
$V_0~\sim~4~\times~10^{-2}~m_P^4$ 
(b). And like the case of the exponential potential while $\alpha$ is dimensional value $m_P^2/16\pi \equiv
1$.}
\end{figure*}

Power-law potentials~\cite{we99} are the potentials like
$$
V(\varphi) = \frac{\lambda{\phi}^n}{n}, \quad n \ge 2.
$$

These potentials are well-studied and they lead to "chaotic inflation"~\cite{linde_ch}. One can
really use them as "inflation part" in the potentials like ones considered by Peebles and 
Vilenkin~\cite{peeb_vil_99}.
Also they 
attract an attention for some their properties~\cite{class,Kolda_Lyth99}. 

The degree of inflationarity for such potentials was studied in our previous 
papers~\cite{we01gr,we01ijmpd,we_new}, so here we will present results only. In the case $n=2$ we have about 
$47\%$
inflation solution in the case of 
field measure and about $63\%$ in the case of angular measure. Increasing $n$ will decrease the 
inflationarity  
in the case of field parametrization and will not change it
for the angular parametrization (\cite{we_new})~-- in the angular measure $63\%$ of all possible 
solutions experience inflation for $n=2,4,6,8$. And in the field measure we have about $30\%$
for $n=4$, $22\%$ for $n=6$ and $17\%$ for $n=8$. 
Damour-Mukhanov potential~\cite{D-M}  
behaves itself like power-law potentials in this sense and the probability of inflation is about $63\%$ 
in the case of the angular 
measure and not less then $47\%$ in the case of the field one (for discussion see~\cite{we_new}).

\subsection{Inverse power-law potentials}

Pioneering studies of the inverse power-law potentials have been done by Ratra and 
Peebles~\cite{peeb_ratr88}. These potentials are like~\cite{ZWS} 
$$
V(\phi) = M^{(4 + q)} \varphi^{-q}.
$$

Potentials of such a kind mostly studied for their scaling properties~\cite{class} and with 
some problems with quintessence~\cite{Kolda_Lyth99}, but also they appear in some SUGRA theories 
as well~\cite{Brax_Martin}.

Here we examine potentials with large enough values of $M$~-- really, for small "observed" 
values~\cite{peeb_vil_99} we obtained that inflationary solutions are not experience enough number of
e-foldings. And it occurs just because of we have a deal with closed model~-- it naturally decrease
number of inflationary solutions.

One of the ways to get large value of $M$ at the inflation stage and small one nowadays is to 
consider $M$ as not an exact constant but as slowly varing (namely decreasing). It would not change
significantly during inflation but from inflation to nowadays can. 

And the results are presented in Fig.~1: panels (a)--(c) represent examples of the initial
condition space where grey point reflects initial condition leading to inflation (in the sense we noted 
above) and black one corresponds to recollapse. These figures are plotted for the case $M~\sim~m_P$
and $q=2$ (a), $q=6$ (b) and $q=8$ (c).
In Fig.~2  we represent the final plot~-- with 
the degree of inflationarity on {\it y} axes and the power $q$ on {\it x} axes. There are three curves: I
corresponds to the case $M~\sim~m_P $, II~-- $M~\sim 0.7~m_P$ and III~-- $M~\sim~0.4~m_P$.
Note that in this case~-- inverse power-law potentials~-- we use angular measure only because in field
measure the space of initial conditions is unbounded.

One can see from Fig.~1 (a)--(c) the areas are asymmetric with respect to $\phi = 0$. It's due to the 
parametrization~-- from (*) one can see that positive $\phi$ corresponds to positive $\dot \varphi$
so we start in this case "down" the potential, at the same time negative $\phi$ corresponds to 
negative $\dot \varphi$ and so we start "up" the potential. Namely due to this effect the area of 
inflationary solutions on the space of initial conditions is so asymmetric.

\subsection{Exponential potential}

Other interesting potential is an exponential one~\cite{peeb_ratr88}
$$
V(\phi) = V_0 \exp ( - \lambda \varphi  ).
$$

Potentials of such a kind naturally appear in high-energy physics in theories with Kaluza-Klein 
compactification, superstrings and supergravity theories~\cite{exp_from_sugra}  
and higher orders 
gravity theories~\cite{exp_from_high_grav}. This potential is also well-known in flat case for its 
tracker property and for some implications for observational cosmology~\cite{ferr_joy98,exp_other,
class}.

Our results are the same for a wide enough range of $V_0~\sim~m_P^4~\div~10^{-10}~m_P^4$. 
So we have only one free 
parameter~-- $\lambda$ and the results are plotted in Fig.~3. 
Example of the area on initial conditions space plotted in Fig.~1(d), where as in the 
previous cases black dot represents pair of initial $(\phi,H)$ not leading to inflation and 
grey dot represents leading one.
Here as well as in the case of inverse power-law potentials we use only angular 
parametrization~-- the initial $\varphi$ is unbounded again.

\subsection{ZWS potential}

\begin{figure*}
\epsfxsize=12cm
\centerline{{\epsfbox{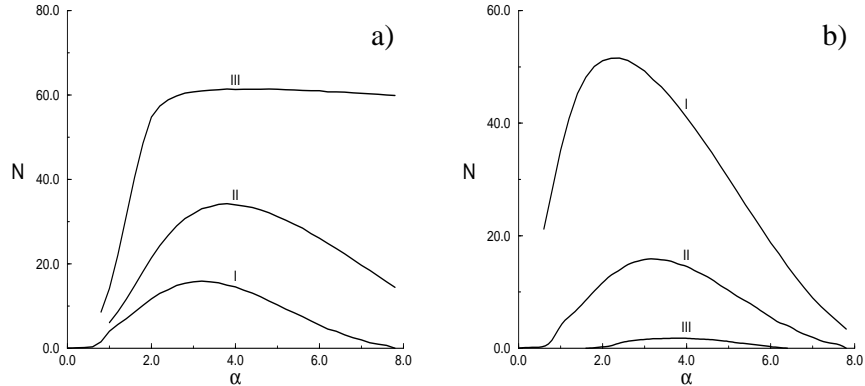}}}
\caption
{ULM potential: the dependence of the degree of 
inflationarity on $\alpha$ for fixed $V_0~=~4~\times~10^{-3}~m_P^4 $ in panel (a) and for fixed 
$\lambda = 0.7~m_P^{-1}$ in panel (b).}

\end{figure*}

This potential
$$
V(\varphi) = V_0 \left[ \exp \left( \frac{\alpha}{\varphi} \right) - 1 \right]
$$

\noindent was suggested by Zlatev, Wang and Steinhardt~\cite{ZWS} while 
exploring scaling solutions, where $V_0$ and $\alpha$ are free parameters.

The dependence of the degree of inflationarity is plotted in Fig.~4: the case for 
$V_0~\sim~4~\times~10^{-4}~m_P^4$ in (a) panel and the case for $V_0~\sim~4~\times~10^{-2}~m_P^4$ in (b). 
Areas of initial 
conditions space leading and not
to inflation look like ones in the case of inverse power-law potential
(Fig.~1(a)--(c)). Due to the potential (runaway) we can use only the trigonometrical measure.

\subsection{ULM potential}

This potential was first considered by L.A. Ure$\tilde {\rm n}$a-L\'opez and 
T. Matos~\cite{Matos_Urena-Lopez00} as a new cosmological tracker 
solution for quintessence. One can say it behaves itself like both exponential and inverse power-law 
potentials:
$$
V(\varphi) = V_0 \sinh^{-\alpha}(\lambda \varphi);
$$
$$
V(\varphi) = V_0 (\lambda \varphi)^{-\alpha} \quad \mbox{for} \; |\lambda \varphi| << 1,
$$
$$
V(\varphi) = 2^{\alpha} \, V_0 e^{-\lambda \varphi \alpha} \quad \mbox{for} \; |\lambda \varphi| >> 1
$$

\noindent so its asymptotic behavior corresponds to an inverse power-law-like potential at early times
and to an exponential one at later times.

The space of initial conditions is like the inverse power-law case (Fig.~1~(a)--(c)), and we
have again only trigonometrical measure. The results for the degree of inflationarity are plotted in Fig. 5
with very degree on {\it y} axes and $\alpha$ on {\it x} axes.  In this case we have two parameters~--
$V_0$ and $\lambda$ and it's easily to understand the influence of them by fixing one of them and
varying another one. Following its in (a) panel we fix 
$V_0~=~4~\times~10^{-3}~m_P^4$ and vary $\lambda$:
$\lambda = 0.7~m_P^{-1}$ (I), $\lambda = 0.5~m_P^{-1}$ 
(II) and $\lambda = 0.21~m_P^{-1}$ (III); in panel (b) we fix $\lambda = 0.7~m_P^{-1}$ and vary $V_0$:
$V_0~=~4~\times~10^{-2}~m_P^4$ (I), $V_0~=~4~\times~10^{-3}~m_P^4$ (II) and 
$V_0~=~4~\times~10^{-4}~m_P^4$ (III).

\subsection{FCDM potential}

This potential can describe both quintessence and
new form of dark matter called~\cite{sahni_wang_99} {\it frustrated cold dark matter} (FCDM) due to 
its ability to frustrate gravitation clustering at small scales. On this way it can help in solving 
some problems with dwarf galaxies halos~\cite{halos}.

Like previous one it have two asymptotic forms:
$$
V(\varphi) = V_0 ( \cosh \lambda \varphi - 1 )^p;
$$
$$
V(\varphi) \simeq \tilde V_0\, e^{p \lambda \varphi} \quad \mbox{for} \; |\lambda \varphi| >> 1 \; 
(\varphi > 0),
$$
$$
V(\varphi) \simeq \tilde V_0 (\lambda \varphi)^{2p} \quad \mbox{for} \; |\lambda \varphi| << 1,
$$

\noindent where $\tilde V_0 = V_0 /2^p$.

\begin{figure*}
\rotatebox{-90}{
\epsfxsize=18cm
\includegraphics{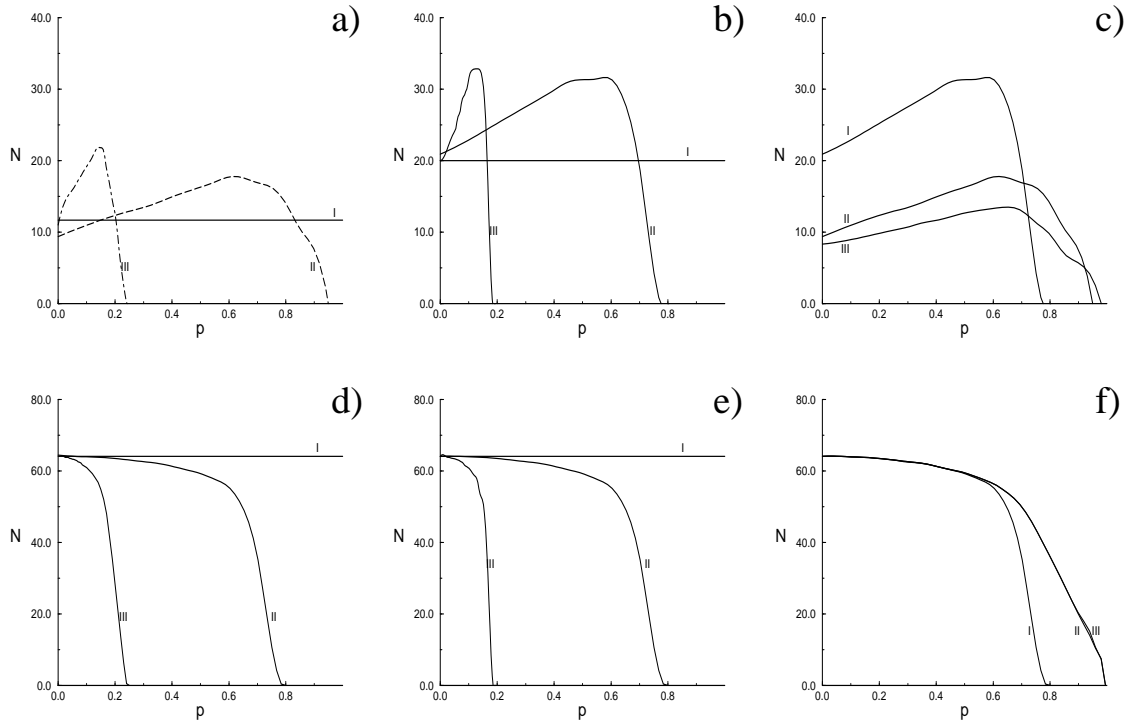}}
\caption
{FCDM potential: the dependence of the degree of inflationarity on $V_0$ and $\lambda$; (a)--(c)
correspond to the case of field parametrization and (d)--(f) correspond to the case of 
trigonometrical parametrization. See text for detailed description of the panels.}
\end{figure*}

In this case like in previous one we have two free parameters (while $p = 0 \div 1$): $V_0$ and 
$\lambda$. The space of 
initial conditions with the area of initial conditions leading to inflation looks like in the case of
inverse power-law (see Fig.~1~(a)--(c)) in the trigonometrical parametrization case and for field
parametrization (enjoy! for this potential we have field parametrization!) examples are presented
in Fig. 1(e) and Fig. 1(f). 
And in Fig.~6 we have presented the dependence curves of the
ratio of inflationary solutions to all possible ones on $p$ ($p \in [0,1]$). The dependence of the
degree of inflationarity 
on both $V_0$ and $\lambda$ is hard to describe so we have presented the influence of $V_0$ and 
$\lambda$ on the curve when one of them is fixed and rest one is variable (as in the previous case). 
As we have noted above, in this case we have
a deal with both field and angular measures. And one can see that the curves behave themselves 
differently in these measures.

Very results are presented in Fig.~6: panels (a)--(c) correspond to the case of field 
parametrization and panels (d)--(f) correspond to the case of trigonometrical parametrization. Let 
us describe them: in Fig. 6(a) and 6(d) there are dependence curves of the degree of inflationarity
on $\lambda$ for $V_0 = 4 \times 10^{-7} m_P$ and curves are: $\lambda = 0.7 m_P^{-1}$ (I), 
$\lambda = 7 m_P^{-1}$ (II) and $\lambda = 30 m_P^{-1}$ (III). In Fig 6(b) and 6(e) we again 
fixed $V_0 = 4 \times 10^{-4} m_P$ and $\lambda = 0.7 m_P^{-1}$ (I), 
$\lambda = 7 m_P^{-1}$ (II) and $\lambda = 30 m_P^{-1}$ (III). To understand the influence of
$V_0$ on the dependence curve one can compare these two cases as well as use Figs. 6(c) and 6(f), 
where we fixed $\lambda = 7 m_P^{-1}$ and curves are $V_0 = 4 \times 10^{-4} m_P$ (I),
$V_0 = 4 \times 10^{-7} m_P$ (II) and $V_0 = 4 \times 10^{-10} m_P$ (III).
One can see that in the limit of small $V_0$ all dependence curves are similair each other.

So from Fig.~6~(a)--(f) one can determine values for $\lambda$ and $V_0$ which correspond to large
enough degree of inflationarity.

\section{Summary and discussions}

In this paper we have investigated a wide class of the quintessence potentials from the point of view 
of the generality of inflation. Some of these potentials are
motivated via particle physics (for review see, for example, \cite{Lyth_Riotto99}) and most of them 
were also studied recently from the point of view of scaling or tracker solutions. So here we have made some 
tests of them~-- are they able to provide our Universe with inflation? And in this way we
obtained the answer~-- yes, closed FRW models filled with the scalar field with these potentials 
experience inflation for a wide enough range of their parameters. So inflation is general for a wide
enough class of the cosmological models.

As we have a deal with closed model, the degree of inflationarity is smaller then in flat (and 
of course open) case~-- at 
the early stage of the Universe's evolution the scale factor was small so presence of nonzero 
curvature might change the situation significantly. But we have seen that even in the case of positive
spatial curvature potentials we have considered behave itself well from this point of view and the
inflation is not suppressed in models filled with the scalar field with these potentials. The 
inflation in closed models was recently studied in the case of pure $\Lambda$-term~\cite{lambda}
and they also have a result that the inflation doesn't suppress.

In this paper we used as determination of inflation the situation when the scale factor increases in
$10^5$ times (about 12 e-foldings) and this approximation works well. For some cases we have checked
its up to $10^{25}$ (about 57 e-foldings) and the results are the same with precision about $0.04\%$. 
So our approximation works well.

\section{Acknowledgments}

We wish to thank N. Savchenko and A. Toporensky for useful discussion.


\begin{thebibliography}{99}

\bibitem{snIa_1}
S.J. Perlmutter {\it et. al.}, Nature (London) {\bf 391}, 51 (1998);
S.J. Perlmutter {\it et. al.}, Astrophys. J. {\bf 517}, 565 (1999).

\bibitem{snIa_2}
A.G. Riess {\it et. al.}, Astron. J. {\bf 116}, 1009 (1998).

\bibitem{CMB_rec}
A. Benoit {\it et al.}, astro-ph/0210306; J.E. Ruhl {\it et al.}, astro-ph/0212229; J.H. Goldstein
{\it et al.}, astro-ph/0212517; D.N. Spergel {\it et al.}, astro-ph/0302209. 

\bibitem{LSS_rec}
M. Colless {\it et al.}, Mon. Not. R. Astron. Soc. {\bf 328}, 1039 (2001); Efslathiou {\it et al.},
Mon. Not. R. Astron. Soc. {\bf 330}, L29 (2002); Verde {\it et al.}, Mon. Not. R. Astron. Soc. 
{\bf 335}, 432 (2002).

\bibitem{als_var00}
V. Sahni and A. Starobinsky, Int. J. Mod. Phys. D {\bf 9}, 373 (2000) [astro-ph/9904398].

\bibitem{var02}
V. Sahni, Class. Quant. Grav. {\bf 19}, 3435 (2002) [astro-ph/0202076].

\bibitem{peeb_ratr02}
P.J.E. Peebles and B. Ratra, Rev. Mod. Phys. (to be published), astro-ph/0207347.

\bibitem{infl}
A.H. Guth, Phys. Rev. D {\bf 23}, 347 (1981); A.D. Linde, Phys. Lett. B {\bf 108}, 389 (1982);
A. Albrecht and P.J. Steinhardt, Phys. Rev. Lett. {\bf 48}, 1220 (1982); Sato, Mon. Not. R. Astron.
Soc. {\bf 195}, 467 (1981); 
A. Starobinsky, Pis'ma A.J. {\bf 4}, 155 (1978), Sov. Astron. Lett. {\bf 4}, 82 (1978).

\bibitem{als-sn-v}
A. Starobinsky {\it et. al.}, Phys. Rev. Lett. {\bf 85}, 2236 (2000); {\bf 85}, 1162 (2000);
A. Starobinsky, JETP Lett. {\bf 68}, 757 (1998).

\bibitem{Linde_rec}
A. Linde, astro-ph/0303245.

\bibitem{four}
V.A. Belinsky, L.P. Grishchuk, Ya.B. Zeldovich and I.M. Khalatnikov,
 JETP {\bf 89}, 346 (1985).

\bibitem{B-Kh}
V.A. Belinsky and I.M. Khalatnikov,  JETP  {\bf 93}, 784 (1987);
V.A. Belinsky, H. Ishihara, I.M. Khalatnikov and H. Sato,
 Prog. Theor. Phys.  {\bf 79}, 676 (1988).

\bibitem{we99}
S.A. Pavluchenko and A.V. Toporensky, Gravitation $\&$ Cosmology {\bf 6}, 241 (2000) [gr-qc/9911039].

\bibitem{we01gr}
S.A. Pavluchenko and A.V. Toporensky, Gravitation $\&$ Cosmology Suppl. {\bf 8}, 168 (2002).

\bibitem{we01ijmpd}
S.A. Pavluchenko, N.Yu. Savchenko and A.V. Toporensky, Int. J. Mod. Phys. 
D {\bf 11}, 805 (2002)  [gr-qc/0111077].

\bibitem{we_new}
S.A. Pavluchenko and A.V. Toporensky, in preparation.

\bibitem{Lyth_Riotto99}
D.H. Lyth and A. Riotto, Phys. Rept. {\bf 314}, 1 (1999) [hep-ph/9807278].

\bibitem{D-M}
T. Damour and V.F. Mukhanov, Phys. Rev. Lett. {\bf 80}, 3440 (1998),
A. Liddle and A.Mazumdar, Phys. Rev. D {\bf 58}, 083508 (1998).

\bibitem{class}
A.R. Liddle and R.J. Scherrer, Phys. Rev. D {\bf 59}, 023509 (1998).


\bibitem{peeb_ratr88}
B. Ratra and P.J.E. Peebles, Phys. Rev. D {\bf 37}, 3406 (1988); P.J.E. Peebles and B. Ratra, 
Astrophys. J., Lett. Ed. {\bf 325}, L17 (1988). 


\bibitem{linde_ch}
A. Linde, Phys. Lett. B {\bf 129}, 177 (1983).

\bibitem{peeb_vil_99}
P.J.E. Peebles and A. Vilenkin, Phys. Rev. D {\bf 59}, 063505 (1999).

\bibitem{sahni_wang_99}
V. Sahni and L. Wang, Phys. Rev. D {\bf 62}, 103517 (2000) [astro-ph/9910097].

\bibitem{halos}
B. Moore {\it et. al.}, astro-ph/9903164; J.F. Navarro and M. Steinmetz, astro-ph/9908114;
S. Ghigna {\it et. al.}, astro-ph/9910166; D.N. Spergel and P.J. Steinhardt, Phys. Rev. Lett. 
{\bf 84}, 3760 (2000); M. Kanionkowski and A.R. Liddle, Phys. Rev. Lett. {\bf 84}, 4525 (2000).

\bibitem{ferr_joy98}
P.G. Ferreira and M. Joyce, Phys. Rev. Lett. {\bf 79}, 4740 (1997); Phys. Rev. D {\bf 58}, 023503 
(1998).

\bibitem{Kolda_Lyth99}
C. Kolda and D.H. Lyth, Phys. Lett. B {\bf 458}, 197 (1999).

\bibitem{Brax_Martin}
Ph. Brax and J. Martin, Phys. Lett. B {\bf 468}, 40 (1999) [astro-ph/9905040]; Phys. Rev. D {\bf 61},
103502 (2000).

\bibitem{exp1}
C. Wetterich, Nucl. Phys. {\bf B252}, 302 (1985); {\bf B252}, 688 (1985); Astron. Astrophys. 
{\bf 301}, 321 (1995); E.J. Copeland {\it et. al.}, Ann. N.Y. Acad. Sci. {\bf 688}, 647 (1993).

\bibitem{exp_from_high_grav}
C. Wetterich, Nucl. Phys. {\bf B252}, 309 (1985); Q. Shafi and C. Wetterich, Phys. Lett. B {\bf 152},
51 (1985); Nucl. Phys. {\bf B289} 787 (1987); J. Halliwell, Nucl. Phys. {\bf B266}, 228 (1986); J.D.
Barrow and S. Cotsakis, Phys. Lett. B {\bf 214}, 515 (1988).

\bibitem{exp_from_sugra}
A.B. Burd and J.D. Barrow, Nucl. Phys. {\bf B308}, 929 (1988); J. Yokoyama and K. Maeda, Phys. Lett. B
{\bf 207}, 31 (1988); P. Bin\'etruy, Phys. Rev. D {\bf 60}, 063502 (1999) [hep-ph/9810553].

\bibitem{exp_other}
E.J. Copeland, A.R. Liddle and D. Wands, Phys. Rev. D {\bf 57}, 4686 (1998);
P.T.P. Viana and A.R. Liddle, Phys. Rev. D {\bf 57}, 674 (1998); T. Barreiro, E.J. Copeland and N.J. 
Nunes, Phys. Rev. D {\bf 61}, 127301 (2000); A. Albrecht and C. Skordis, Phys. Rev. Lett. {\bf 84},
2076 (2000).

\bibitem{ZWS}
I. Zlatev, L. Wang and P.J. Stainhardt, Phys. Rev. Lett {\bf 82}, 896 (1999); Phys. Rev. D {\bf 59}, 
123504 (1999).

\bibitem{Matos_Urena-Lopez00}
L.A. Ure$\tilde {\rm n}$a-L\'opez and T. Matos, Phys. Rev. D {\bf 62}, 081302(R) (2000).

\bibitem{lambda}
J.P. Uzan, U. Kirchner and G.F.R. Ellis, [astro-ph/0302597].



\end{thebibliography}
\end{document}